**All-nitrogen cages and molecular crystals: Topological rules, stability, and pyrolysis paths**

Konstantin P. Katin[1,2], Valeriy B. Merinov[1], Alexey I. Kochaev[3], Savas Kaya[4], Mikhail M. Maslov[1,2,*]

[1]Department of Condensed Matter Physics, National Research Nuclear University "MEPhI", Kashirskoe Sh. 31, Moscow, 115409, Russian Federation

[2]Laboratory of Computational Design of Nanostructures, Nanodevices, and Nanotechnologies, Research Institute for the Development of Scientific and Educational Potential of Youth, Aviatorov Str. 14/55, Moscow, 119620, Russia

[3]Ulyanovsk State University, 42 Leo Tolstoy str., Ulyanovsk, 432017, Russia

[4]Faculty of Science, Department of Chemistry, Cumhuriyet University, Sivas, Turkey

*Mike.Maslov@gmail.com*

ABSTRACT

We have combined *ab initio* molecular dynamics with the intrinsic reaction coordinate in order to investigate the mechanisms of stability and pyrolysis of $N_4 \div N_{120}$ fullerene-like nitrogen cages. The stability of the cages was evaluated in terms of the activation barriers and the activation Gibbs energies of their thermal-induced breaking. We found that binding energies, bond lengths, and quantum-mechanical descriptors failed to predict the stability of the cages. However, we derived a simple topological rule that adjacent hexagons on the cage surface resulted in its instability. For this reason, the number of stable nitrogen cages is significantly restricted in comparison with their carbon counterparts. As a rule, smaller clusters are more stable, whereas earlier proposed rather large cages collapse at room temperature. The most stable all-nitrogen cages are $N_4$ and $N_6$ clusters, which can form the van-der-Waals crystals with the densities of 1.23 and 1.36 g/cm$^3$, respectively. Examination of their band structures and densities of electronic states shows that they are both insulators. Their power and sensitivity are not inferior to the modern advanced high-energy nanosystems.

**Introduction**

The formation of the $N_2$ molecule with the triple N≡N bond from two isolated nitrogen atoms results in the release of a large amount of energy (~ 10 eV [1]). Thus, nitrogen is considered as the primary component of the most high-energy compounds. Such compounds usually contain also carbon atoms that provide stability of the whole molecule framework decorated by oxygen-containing groups or other oxidants. RDX, HMX, and CL-20 are well-known examples of traditional high-energy structures with a carbon-nitrogen frame. In recent years, more advanced energy materials have been proposed with a similar carbon-nitrogen architecture [2; 3; 4].

The recent synthesis of non-molecular nitrogen [5; 6] has renewed research interest in all-nitrogen structures without carbon additives. Compared to carbon-nitrogen compounds, pristine nitrogen systems are more powerful and safer for the environment. In particular, closed full-nitrogen $N_n$ cages with three-coordinated atoms and single N–N bonds are very attractive and environmentally friendly oxygen-free fuels with an extremely high energy capacity. The transformation of single N–N bonds into the triple bond is a very exothermic process. The breakdown of $N_n$ cages into $N_2$ molecules gives more than 50 kcal/mol per nitrogen atom [7]. However, in practice, $N_n$ cages possess low stability and have not yet been synthesized and measured. Unlike carbon fullerenes, fully nitrogen cages do not retain their structure under ambient conditions, since they are energetically unfeasible compared to $N_2$ molecules. In addition, the activation energies of their decay processes are rather low. Thus, most $N_n$ cages can be prepared and stabilized only at extremely high pressures, similar to recently obtained forms of non-molecular nitrogen [5; 6]. The introduction of carbon atoms, oxygen, or other atoms into the nitrogen cages leads to more stable structures [3; 4; 8], which, however, are not as efficient and environmentally friendly as all-nitrogen hypothetical analogs [7; 9].

Many researchers have looked at the structures of $N_n$ cages to find some that are stable enough to be used in practice. All-nitrogen $N_4$ [10], $N_6$ [11], $N_8$ [12], $N_{10}$ [9], $N_{12}$ [9; 13], $N_{14}$ [14], $N_{16}$ [14], $N_{18}$ [15], $N_{20}$ [9; 16], $N_{24}$ [17], $N_{30}$ [17], and $N_{36}$ [17] structures turned out to be local energy minima in the systems' configuration spaces. A genetic algorithm was recently applied to search for configurations of global minima of $N_n$ ($n = 3 \div 10$) clusters [18]. Among huge $N_n$ cages with $n > 36$, only a limited number of systems were considered. All of them are tubular clusters similar to nanotubes [19; 20; 21]. Unexpectedly, the $N_{60}$ cluster with the $I_h$ symmetry (the nitrogen analog of the traditional $C_{60}$ fullerene) was found to be unstable even at zero temperature [22]. Low kinetic stability can be considered as the Achilles heel of all previously studied systems, even if they were true energy minima in the configuration space.

In the presented paper, we carry out a systematic study of the structure and stability of $N_n$ fullerene-like cages. Our goal is to identify the general relationships between topology and stability of nitrogen buckyballs. Unlike previous calculations, we focused on the pyrolysis mechanisms and the corresponding activation barriers, rather than on the relative formation energies of the structures under consideration. We identified the most stable $N_n$ cages, explained the reasons for their stability, and investigated the possibilities of their crystallization.

**Computational Details**

Finite-size nitrogen cages were simulated in the frame of the density functional theory with the B3LYP functional [23; 24] and 6-311G(d,p) basis set [25]. GAMESS software [26] was used for optimizing stable and transition states, calculations of molecules' vibrational eigenmodes, and following along the intrinsic reaction coordinates. Default numerical thresholds were applied in all cases. Thermodynamic quantities (Gibbs energy and enthalpy) were calculated from molecules' eigenfrequencies in the frame of ideal gas approximation. Binding energies $E_b$ of $N_n$ cages concerning molecular nitrogen were calculated as

$$E_b = \frac{1}{n}\left(E(N_n \text{ cage}) - 0.5nE(N_2 \text{ molecule})\right).$$

Activation Gibbs energy $\Delta G$, activation enthalpy $\Delta H$, and energy barrier $\Delta U$ were obtained as the differences of corresponding values for transition and initial configurations of nitrogen cages. The lifetime of a molecule $t$ before the thermal decomposition at a temperature $T$ was estimated using the Arrhenius law

$$t = (Ag)^{-1} \exp\left(-\frac{\Delta U}{kT}\right),$$

where $k$ is the Boltzmann constant. Frequency factors $A$ were defined in accordance with the Vineyard's rule [27]

$$A = \frac{\prod_{i=1}^{3n-6} \omega}{\prod_{i=1}^{3n-7} \omega'},$$

where $\omega$ and $\omega'$ are real eigenfrequencies at stable and transition states, respectively. Note that the imaginary eigenfrequency of the transition state is not included in the multiplication. Degeneracy factor $g$ is equal to the number of equivalent decay paths ($g > 1$ for high-symmetry $N_n$ cages).

*Ab initio* molecular dynamics was carried out using a high-performance GPU-based TeraChem package [28; 29; 30; 31] with the same B3LYP/6-311G(d,p) level of theory. The initial atomic displacements and velocities were generated in accordance with the desired temperature. During the simulation, the given temperature was maintained by the Langevin thermostat [32]. The time step for all simulations was 0.1 fs, which was sufficient to take into account all molecular vibrations correctly. Systems' dynamical trajectories were visualized using the ChemCraft program [33].

To study geometry and electronic structure properties of crystals that consist of the non-molecular nitrogen systems, we used another density functional theory approach and its implementation in the QUANTUM Espresso 6.5 program package [34; 35]. The generalized gradient approximation (GGA) in the Perdew-Burke-Ernzerhof (PBE) functional form for the exchange-correlation energy [36], and the projector-augmented-wave (PAW) method for the electron-ion interaction [37; 38] was used to perform the calculations. The value of cut-off energy for the plane-wave basis set was chosen as 120 Ry (1632 eV). The weak van der Waals interactions between the non-covalently bound nitrogen atoms are taken into account by using the D3 Grimme dispersion corrections [39]. The geometry optimization of the nitrogen crystal was performed without any symmetry constraints until the Hellman-Feynman forces acting on the atoms became smaller than $10^{-6}$ hartree/bohr. Supercell parameters were also optimized. The Brillouin zone integrations were performed by using the Monkhorst-Pack k-point sampling scheme [40] with the 8×8×8 mesh grid. For the non-self-consistent field calculations, the *k*-point grid size of 24×24×24 was used. The Methfessel-Paxton smearing [41] was used for the geometry relaxation with the smearing width of 0.02 eV. Still, for the calculations of the electronic density of states, the Böchl tetrahedron method [42] was employed. The electronic structure properties were elucidated through the analysis of the sample band structure and its electronic density of states.

**Results and Discussion**

**A. The first insight on the bicyclic hydro-nitrogen molecules**

Hypothetical bicyclic hydro-nitrogen molecules are the smallest possible systems containing adjacent nitrogen rings with the single N–N bonds. We constructed $N_{10}H_8$, $N_9H_7$, $N_8H_6$, $N_7H_5$, and $N_6H_4$ naphthalene-like molecules with hexagon/hexagon, hexagon/pentagon, pentagon/pentagon, pentagon/square, and square/square interfaces, respectively (see Fig. 1a). Unlike their carbon counterparts, these molecules did not remain flat during geometry optimization.

We found that the hexagon/pentagon and the pentagon/pentagon were the only stable configurations with positive eigenfrequencies. Other structures are not true energy minima. Potential energy paths corresponding to the breaking of joint hexagon/pentagon and pentagon/pentagon bonds are shown in Fig. 1b and 1c (the energy barriers are 0.58 and 0.47 eV, respectively). Therefore, the pentagon/pentagon interface seems to be the most suitable. The hexagon/pentagon interface possesses lower stability, while the adjacent hexagons lead to the complete instability. Note that the carbon structures exhibit the opposite behavior, avoiding adjacent pentagons according to Kroto's isolated pentagon rule [43].

However, bicyclic molecules can only give qualitative insight into the stability of all-nitrogen cages. They do not reproduce local curvature effects, which is especially crucial for small cages, and other effects such as three-dimensional aromaticity. Moreover, hydrogen passivation distorts the charge distribution in the nitrogen rings. Thus, the conclusions made from the analysis of bicyclic nitrogen-containing molecules should be verified on more suitable all-nitrogen systems.

### B. Nitrogen cages with the adjacent hexagons

As a next step, we constructed nitrogen analogs of low-energy isomers of carbon fullerenes from $C_{20}$ to $C_{120}$, created using the FULLERENE 4.5 code [44]. All possible isomers $C_{20} - C_{32}$, as well as all low-energy isomers $C_{34} - C_{120}$, included in the Tomanek database [45] were used as the initial sample geometry for all-nitrogen cages. Note that the typical length of a single N–N bond (1.45 Å) is close to the lengths of C–C bonds in the fullerenes under consideration. However, all the corresponding $N_n$ systems lose their fullerene-like shape when the geometry is optimized. The only exceptions were the clusters $N_{20}$ and $N_{24}$, which did not contain adjacent hexagons. For all other cages, optimization leads to the breaking of those N–N bonds that are common for the neighboring hexagons, and the subsequent destruction of the nitrogen cage. Thus, we conclude that the presence of adjacent hexagons leads to cage instability.

The only known $N_n$ cages with the adjacent hexagons are tubular structures, first considered in Refs. [19; 20; 21]. The structure of the first members belonging to $N_n$ tubular systems is shown in Fig. 2a. The extremely small diameters of these tubes provide stability due to the curvature effects and concave hexagon shapes. Note that the large diameter $N_n$ tubular structures are unstable [46]. We test the stability of the tubular systems shown in Fig.2 using *ab initio* molecular dynamics (AIMD) at $T = 700$ K for one picosecond. We observed their decomposition into $N_2$ molecules, and short $N_n$ chains ($n = 3 \div 7$) started with breaking the adjacent sides of the neighboring hexagons. We conclude that their stability is too low even for their identification. Such nitrogen nanostructures can hardly be used for any practical applications under ambient conditions. Our calculations confirm that, unlike carbon, nitrogen avoids the formation of adjacent hexagonal rings. Note that at the moment, compounds containing adjacent $N_6$ rings have not yet been experimentally synthesized.

### C. Nitrogen cages with the isolated hexagons

A large number of hexagons on the fullerene surface leads to a small curvature and large size of the system. To avoid adjacent pentagons, carbon clusters $C_n$ should be sufficiently large ($n$

≥ 60). In contrast, the hexagons on the surface of nitrogen clusters should be separated by pentagons or smaller polygons, and therefore the fracture of the hexagons should be low. Thus, stable nitrogen clusters $N_n$ should have a significant curvature and small size ($n \leq 24$). Inspired by small carbon fullerenes and fullerene-like cages (including tetrahedranes, prismanes, non-classical fullerenes, etc.), we have constructed the corresponding nitrogen structures. All the nitrogen cages satisfying the isolated hexagon rule that are true local energy minima are presented in Fig. 3. The kinetic stability of these systems was investigated using AIMD at $T = 700$ K for one picosecond. Only the five smallest nitrogen clusters, labeled as (a) - (e) in Fig. 3, conserved their identity during the AIMD simulation (their pyrolysis mechanisms will be described in more detail in the next subsection). Larger clusters are decomposed. Therefore, they should be considered as low-stable systems. Note that N–N bonds in all truly stable and low-stable cages are longer than 1.4 Å, indicating that the bond type is single. The maximum lengths of N–N bonds, binding energies $E_b$, and HOMO-LUMO gaps for all cages are shown in Fig. 4(a)-(c). It should be noted that all these characteristics (and their values) do not correlate with the kinetic stability and, therefore, cannot be used as indicators of stable systems. Quantum-chemical descriptors calculated from the energies of the frontiers orbitals are shown in Fig. S1 in Supplementary Materials. None of the descriptors show a direct correlation with the kinetic stability of the corresponding nitrogen cage.

To get a complete understanding of the origin of the stability of nitrogen cages, we analyzed the mechanisms of their pyrolysis. We adopted the reaction mechanism from the *ab initio* molecular dynamics. Knowing the reaction mechanism, we perform an intrinsic reaction coordinates calculation. The preliminary transient geometries obtained from the AIMD simulations are further optimized to find the true transient configurations (the corresponding atomic coordinates are collected in Supplementary Materials). Energy barriers $U$ are shown in Fig. 4d. One can define two distinct groups of clusters: "stable" with $U > 0.5$ eV and "low-stable" characterized by the value of $U < 0.25$ eV. The analysis of the energy barriers confirms the previous AIMD simulation, where the same stable clusters were distinguished.

### D. Pyrolysis of the most stable nitrogen cages and the possibility of further stabilization

Pyrolysis of five "stable" nitrogen cages is analyzed in more detail. Rate-dependent reaction steps, the corresponding transition states, and products are shown in Fig. 5. Thermodynamic descriptors (activation enthalpies and Gibbs activation energies) at various temperatures are collected in Table 1. These descriptors weakly depend on temperature for all clusters, since the initial and transition configurations are not so different, and their vibrational energies and entropies slightly differ from each other.

According to Table 1, only $N_4$ and $N_6$ clusters have a reasonable lifetime at room temperature. Thus, they should be considered as the most promising high-energy-density structures. More extensive systems need additional stabilization with low temperature or high pressure. According to our previous simulations, small nitrogen clusters can be substantially stabilized via spatial constraints [47] (for example, inside porous materials or carbon cages [48]). Due to the compact shape of small nitrogen clusters, spatial confinement can increase their decomposition activation barrier by 0.5 eV or more [48] that is sufficient to stabilize them at room temperature but hardly useful at higher temperatures.

### E. Van der Waals nitrogen crystals

Recently, some crystal structures have appeared based on the chains $N_6$ [49] and $N_8$ [50], as well as other single-bonded all-nitrogen systems [51; 52]. According to the above calculations, the $N_4$ and $N_6$ clusters are very stable and can be considered as an alternative to nitrogen chains. Their activation barriers are even higher than the corresponding values for experimentally observed highly strained hydrocarbons (for example, tetrahedrane $C_4H_4$ [53] or polymethylcubanes $C_8H_q(CH_3)_{8-q}$ [54]). Thus, $N_4$ and $N_6$ are the most suitable candidates for building blocks for the high-energy-density crystals.

There are many algorithms for predicting the crystal structure. The most famous of them are USPEX [55] and CALYPSO [56]. However, they seem rather excessive for the construction of relatively simple molecular crystals in which the $N_4$ and $N_6$ cages conserve their shape. Isolated nitrogen cages bind in such crystals due to the weak Van der Waals interaction. Instead, we constructed three initial lattices for each crystal (sc, bcc, and fcc) and optimized atomic positions along with lattice vectors without any symmetry constraints. Then the crystals with the lowest potential energy were considered. Their crystallographic parameters are summarized in Table 2. Their band structures and electron densities of states are shown in Fig. 6. Coordinates of the atomic positions are available in Supplementary Materials.

Assuming the detonation of these crystals with the release of gaseous $N_2$, we estimated the heat of detonation $Q$ as the difference in energies between crystals and nitrogen molecules. In addition, the characteristics of detonation (detonation velocity $D$, adiabatic exponent $\Gamma$, detonation pressure $P$) can be estimated from $Q$ and crystal density $p$ in accordance with the Xiong empirical rules [57] that, in our case, have the following forms:

$$D\left[\frac{m}{s}\right] = 67.6\left(Q\left[\frac{cal}{g}\right]\right)^{0.5} + 3300.2\, p\left[\frac{g}{cm^3}\right],$$

$$\Gamma = 1.25 + 3.8\left(1 - \exp\left(-0.546\, p\left[\frac{g}{cm^3}\right]\right)\right),$$

$$P[\text{kbar}] = p\left[\frac{g}{cm^3}\right]\left(D\left[\frac{m}{s}\right]\right)^2 \cdot 10^{-5}/(1+\Gamma).$$

The calculation results are also shown in Table 2. It can be seen that the proposed all-nitrogen crystals have detonation parameters comparable to conventional high-energy materials.

It is more difficult to predict the impact sensitivity of high-energy crystals, which can be characterized by the typical parameter $h_{50\%}$ [58]. Many approaches have been proposed based on the frontier orbitals, electrostatic potentials, and other electronic characteristics [58; 59]. However, their accuracy for the novel compounds is rather questionable [60]. Rice and Hare have indicated that determining the $h_{50\%}$ based on the $Q$ value is preferable [60]. Here we have adopted the following model [59]:

$$h_{50\%}[\text{cm}] = 27.8 + 0.11\exp\left(-11.08\left(Q\left[\frac{\text{kcal}}{g}\right] - 1.66\right)\right)$$

High $Q$ values result in a negligibly small second term, so the formula gives a satisfactory $h_{50\%}$ value of 27.8 cm for both crystals (see Table 2), which is two times higher than the corresponding value for ε-CL-20 (12 cm [61]). Note that this is only a qualitative estimation of impact sensitivity. However, this confirms the potential applicability of the nitrogen crystals under consideration.

CONCLUSION

In this paper, we analyzed the structure and stability of all-nitrogen fullerene-like clusters and molecular crystals based on them. According to the formulated topological rule, confirmed by density functional theory calculations, only small nitrogen cages with single N-N bonds are stable. In terms of stability, they can compete with actively studied nitrogen chains. An increase in cage size always leads to a sharp decrease in its stability. Thus, the range of nitrogen cages is significantly limited. Larger structures can be formed from small building blocks due to non-covalent interactions.

However, our conclusions are valid only for convex cages. We admit that the concave shape of the nitrogen structure can contribute to its stability. However, the curvature of such a hypothetical construction must change its shape many times in order to avoid large convex regions. The study of such astralene-like structures is beyond the scope of this study. We believe that the topological rules presented here will facilitate the further search for new clusters and crystals of nitrogen.

Like strained hydrocarbon frameworks, small nitrogen clusters are stable enough to form molecular crystals. The performed analysis of the two most promising all-nitrogen structures confirmed their applicability as competitive, high-energy materials.

ACKNOWLEDGMENTS

The reported study was funded by RFBR according to the research project No. 18-32-20139 mol_a_ved.

Table 1. Characteristics of the decay of "stable" nitrogen cages $N_n$: energy barriers $U$, degeneracy factors $g$, imaginary frequencies of the transition state $\omega$, activation enthalpies $\Delta H$, activation Gibbs energies $\Delta G$, and mean lifetimes $t$ at different temperatures.

|  | $U$ (eV) | $g$ | $\omega$ (cm$^{-1}$) | $\Delta H$ (kJ/mol) | | | $\Delta G$ (kJ/mol) | | | $t$ (s) | | |
|---|---|---|---|---|---|---|---|---|---|---|---|---|
|  |  |  |  | 300 K | 400 K | 500 K | 300 K | 400 K | 500 K | 300 K | 400 K | 500 K |
| N$_4$ (a) | 2.46 | 4 | 664.1$i$ | 230.5 | 230.5 | 226.1 | 229.1 | 228.7 | 228.3 | 9.6·10$^{26}$ | 4.6·10$^{16}$ | 2.9·10$^{10}$ |
| N$_6$ (b) | 1.45 | 2 | 721.1$i$ | 132.0 | 132.1 | 132.0 | 130.7 | 130.2 | 129.8 | 1.6·10$^{10}$ | 1.3·10$^{4}$ | 3.0 |
| N$_8$ (c) | 0.85 | 6 | 1658.4$i$ | 70.6 | 71.4 | 71.7 | 65.8 | 64.1 | 62.3 | 6.1·10$^{-2}$ | 1.6·10$^{-5}$ | 1.2·10$^{-7}$ |
| N$_8$ (d) | 0.69 | 4 | 746.4$i$ | 58.7 | 59.0 | 59.0 | 55.1 | 53.9 | 52.6 | 5.1·10$^{-4}$ | 6.6·10$^{-7}$ | 1.2·10$^{-8}$ |
| N$_{10}$ (e) | 0.56 | 10 | 566.7$i$ | 47.8 | 47.9 | 47.6 | 46.1 | 45.5 | 45.0 | 4.3·10$^{-6}$ | 1.8·10$^{-8}$ | 6.9·10$^{-10}$ |

Table 2. Characteristics of the N$_4$ and N$_6$ molecular crystals: crystallographic parameters, density $p$, the heat of detonation $Q$, detonation velocity $D$, adiabatic exponent $\Gamma$, detonation pressure $P$, and typical impact sensitive parameter $h_{50\%}$.

|  | Crystallography data | $p$ (g/cm$^3$) | $Q$ (cal/g) | $D$ (m/s) | $\Gamma$ | $P$ (Kbar) | $h_{50\%}$ (cm) |
|---|---|---|---|---|---|---|---|
| N$_4$ | $a = 5.3$ Å, $b = c = 4.6$ Å<br>$\alpha = 70°$, $\beta = \gamma = 55°$ | 1.23 | 2679.3 | 7550 | 3.1 | 170.4 | 28.7 |
| N$_6$ | $a = 5.7$ Å, $b = 4.8$ Å, $c = 6.0$ Å<br>$\alpha = 67.5°$, $\beta = 50.6°$, $\gamma = 54.4°$ | 1.36 | 3219.1 | 8310 | 3.2 | 220.9 | 28.7 |

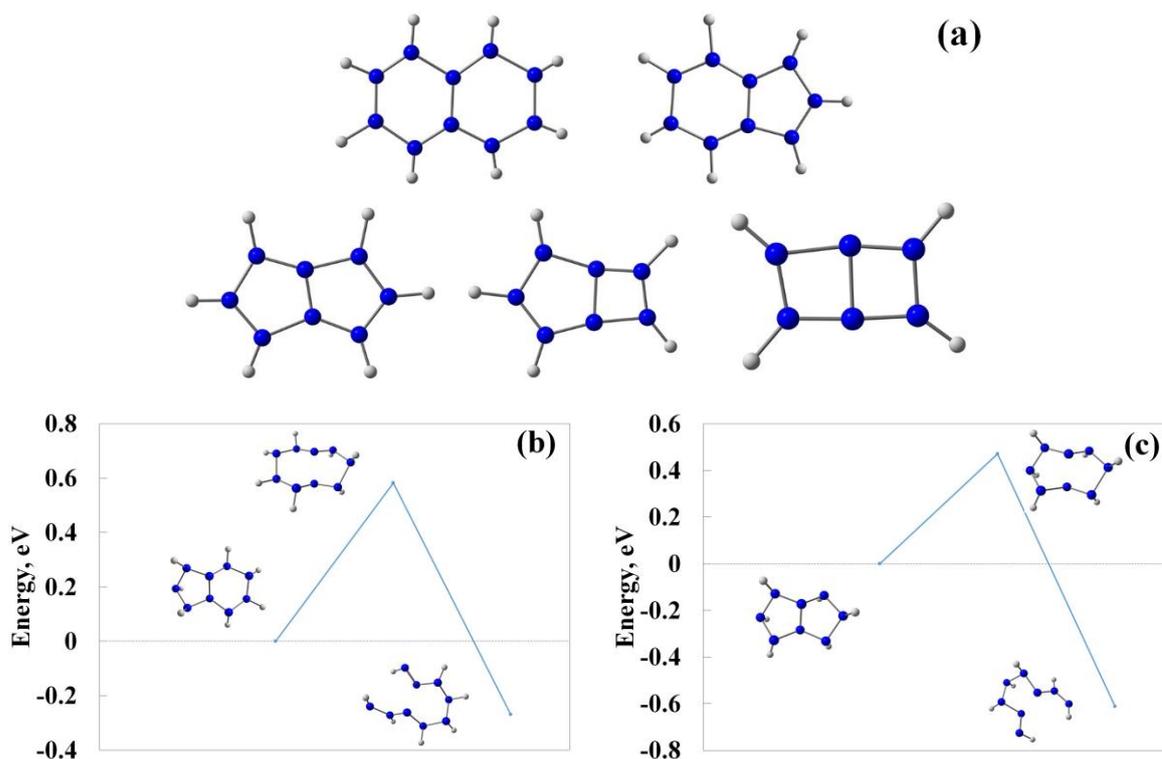

**Fig. 1.** Hypothetical bicyclic hydro-nitrogen molecules $N_{10}H_8$, $N_9H_7$, $N_8H_6$, $N_7H_5$, and $N_6H_4$ (a). Energy diagram of the N-N bond rupture, which is typical for adjacent rings (b)-(c).

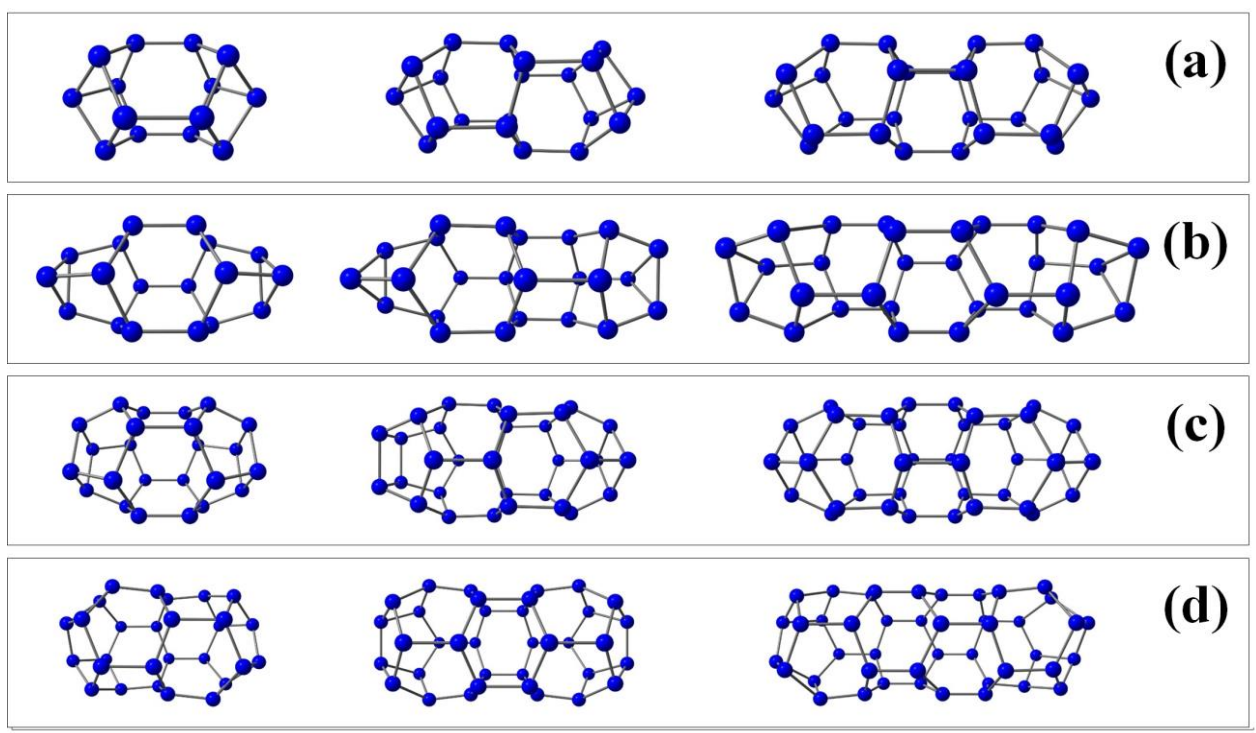

**Fig. 2.** Examples of tubular nitrogen clusters with adjacent hexagons. Nitrogen nanotubes with wall surface formed by three hexagons that contain caps at the ends closed by one (a) and three (b) nitrogen atoms. Nitrogen nanotubes with wall surface formed by four hexagons that contain caps at the ends closed by four (c) and two (d) nitrogen atoms.

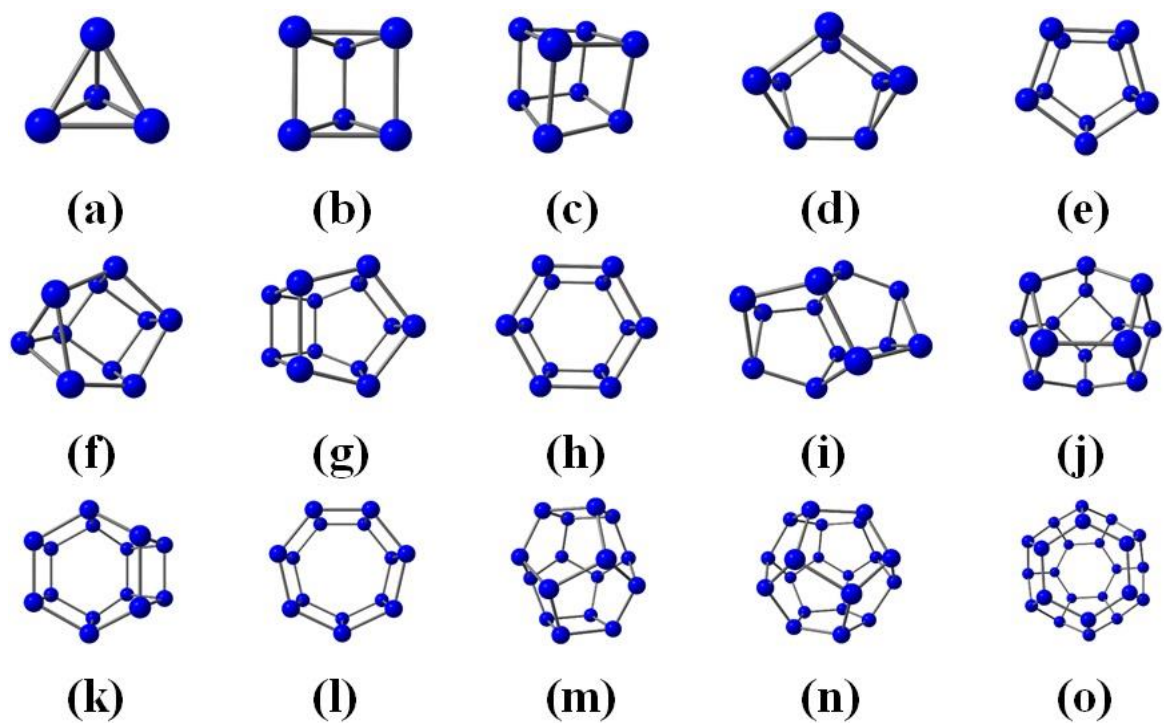

**Fig. 3**. Nitrogen cage structures with isolated hexagons. Atomic structures (a)-(e) are stable, while other systems possess low stability.

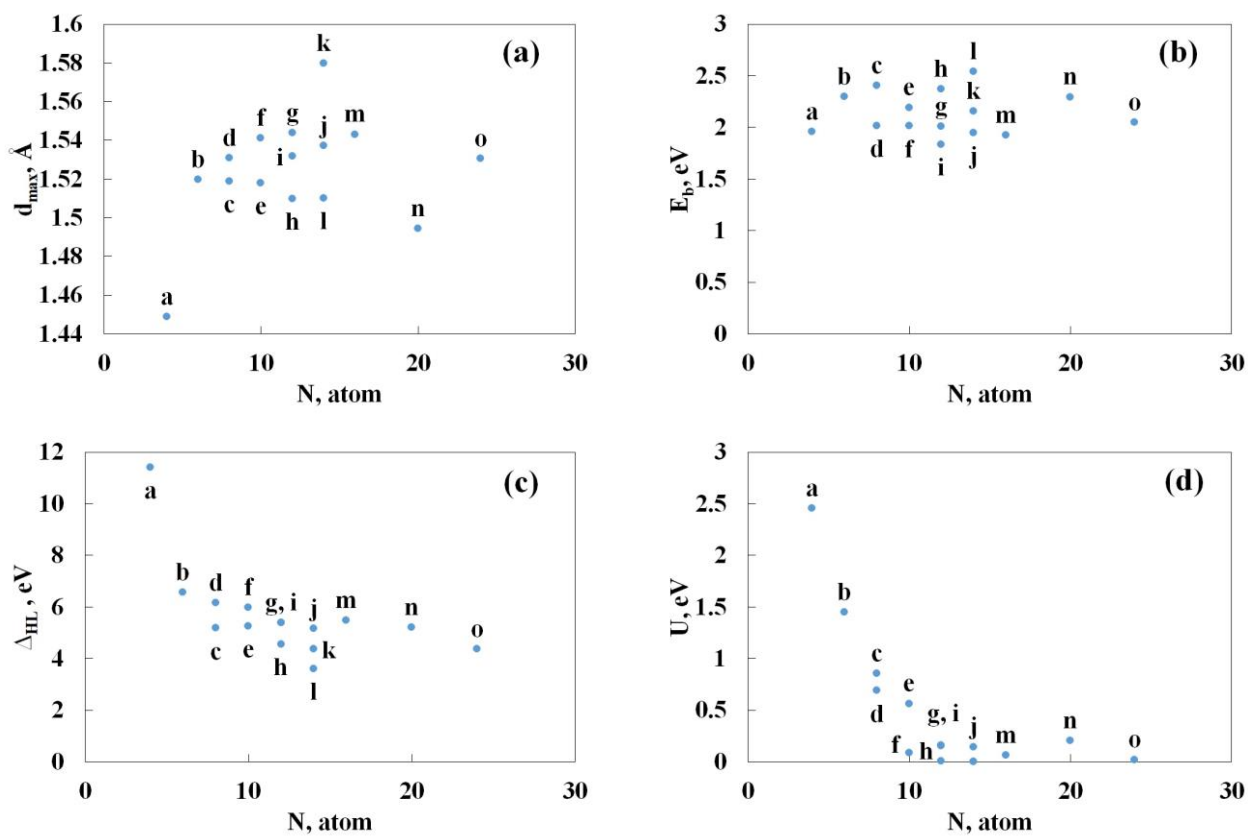

**Fig. 4**. The maximum lengths of N–N bonds (a), binding energies $E_b$ (b), HOMO-LUMO gaps (c), and activation barriers $U$ (d) for corresponding nitrogen cages presented in Fig. 3. Letter designations correspond to designations in Fig. 3.

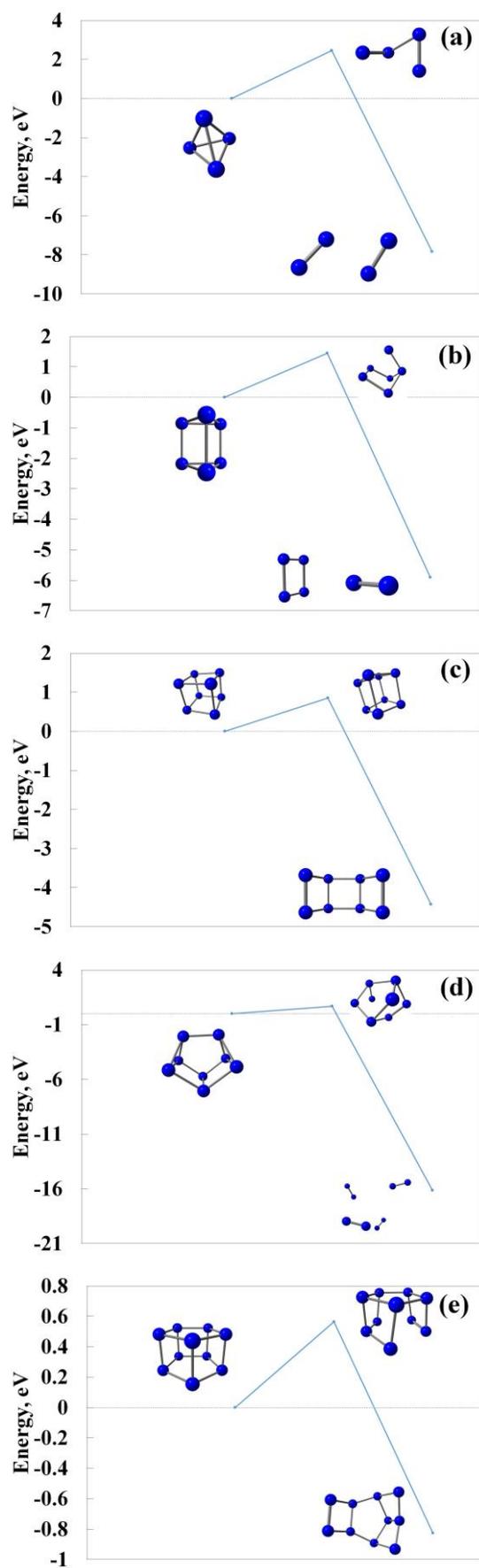

**Fig. 5.** Pyrolysis mechanisms of stable nitrogen clusters $N_4$ (a), $N_6$ (b), $N_8$ (c), $N_8$ (d), and $N_{10}$ (e).

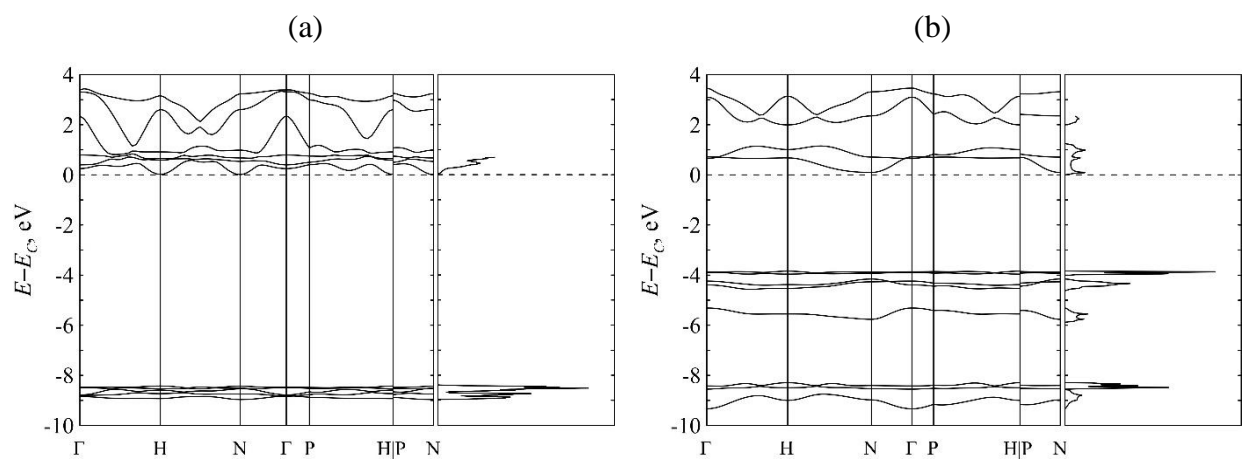

**Fig. 6.** Band structures (left) and electron densities of states (right) for the most stable molecular crystals $N_4$ (a) and $N_6$ (b). Energy $E_C$ denotes the bottom of the conduction band.

TOC Graphics

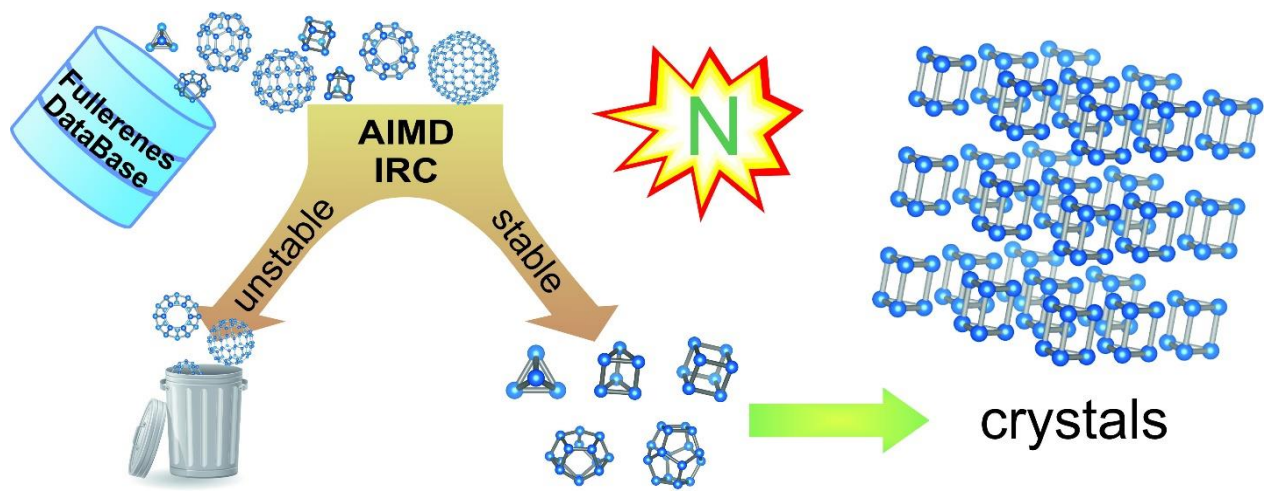

All-nitrogen fullerenes and strained cages were studied using ab initio molecular dynamics, and new high-energy materials were proposed.